\documentclass[aps,prd,longbibliography,reprint,twocolumn,amsmath,amssymb,amsfonts,showpacs,footnote]{revtex4-1}

\usepackage{ifpdf}
\usepackage{dcolumn}
\usepackage{enumitem}
\usepackage{bm}
\usepackage{here}
\usepackage{float}
\floatstyle{plaintop}
\restylefloat{table}
\ifpdf

      \usepackage[pdftex]{graphicx}  
      \usepackage[pdftex]{epsfig}

      \usepackage[pdftex]{hyperref}
\hypersetup
	{
		colorlinks,%
		citecolor=blue,%
		linkcolor=red,%
		urlcolor=blue,%
	}

\else

      \usepackage[dvips]{graphicx}  
      \usepackage[dvips]{epsfig}

      \usepackage[dvipdfm]{hyperref}
      \hypersetup
	{
		colorlinks,%
		citecolor=blue,%
		linkcolor=red,%
		urlcolor=blue,%
	}

\fi

\begin{document}

\title{Electromagnetic radiation generated by a charged particle plunging into a Schwarzschild black hole: Multipolar waveforms and ringdowns}

\author{Antoine Folacci}
\email{folacci@univ-corse.fr}
\affiliation{Equipe Physique
Th\'eorique, \\ SPE, UMR 6134 du CNRS
et de l'Universit\'e de Corse,\\
Universit\'e de Corse, Facult\'e des Sciences, BP 52, F-20250 Corte,
France}

\author{Mohamed \surname{Ould~El~Hadj}}
\email{med.ouldelhadj@gmail.com}
\affiliation{Equipe Physique
Th\'eorique, \\ SPE, UMR 6134 du CNRS
et de l'Universit\'e de Corse,\\
Universit\'e de Corse, Facult\'e des Sciences, BP 52, F-20250 Corte,
France}

\date{\today}

\begin{abstract}

Electromagnetic radiation emitted by a charged particle plunging from slightly below the innermost stable circular orbit into a Schwarzschild black hole is examined. Both even- and odd-parity electromagnetic perturbations are considered. They are described by using gauge invariant master functions and the regularized multipolar waveforms as well as their unregularized counterparts constructed from the quasinormal-mode spectrum are obtained for arbitrary directions of observation and, in particular, outside the orbital plane of the plunging particle. They are in excellent agreement and the results especially emphasize the impact of higher harmonics on the distortion of the waveforms.

\end{abstract}


\maketitle

\tableofcontents

\section{Introduction}
\label{Intr}

In this article, we theoretically construct and numerically obtain the regularized multipolar electromagnetic waveform generated by an electric charge plunging from slightly below the innermost stable circular orbit (ISCO) into a Schwarzschild black hole (BH) and analyze its late-stage ringdown phase in terms of quasinormal modes (QNMs). This is achieved for both even- and odd-parity electromagnetic perturbations. Analogous problems concerning the excitation of a BH by an electric charge and the generation of the associated electromagnetic radiation have been considered since the beginning of the seventies (for pioneering works on this subject, see the lectures by Ruffini \cite{Ruffini:1973pta} in Ref.~\cite{DeWitt:1973uma} and references therein). Currently, even if they oversimplify the accretion of charged matter by a BH or the coalescence of charged stars and BHs \cite{Cardoso:2003cn}, such problems could nevertheless be of great interest with the emergence of multimessenger astronomy which combines the detection and analysis of gravitational waves with those of other types of radiation for a better understanding of our ``violent Universe". In particular, it is interesting to study the electromagnetic partner of the gravitational signal during accretion of charged fluids onto black holes \cite{Degollado:2014dfa,Moreno:2016urq} with in mind the possibility to test the BH hypothesis (for a recent review on this subject, see Ref.~\cite{Bambi:2015kza}). It should also be noted that, even if lots of problems concerning energy extraction from charged particles falling into BHs have been until now considered , the case we intend to examine here has never before been discussed. This is rather surprising: indeed, the ISCO as the last stable orbit for massive particles plays a central role in the accretion process \cite{Abramowicz:2011xu}.

Our paper is organized as follows. In Sec.~\ref{SecII}, after a brief presentation of electromagnetism in the Schwarzschild spacetime, we establish theoretically the expression of the waveforms emitted by a charged point particle plunging from the ISCO into the BH. Here, we work in the frequency domain and use the standard Green's function techniques. Moreover, we consider both the even- and odd-parity electromagnetic perturbations for arbitrary $(\ell,m)$ modes and work with the gauge invariant master functions introduced by Ruffini, Tiomno and Vishveshwara in Ref.~\cite{Ruffini:1972pw} (see also Refs.~\cite{Cunningham:1978zfa,Cunningham:1979px}) which are solutions of the Regge-Wheeler equation with source term. In Sec.~\ref{SecIII}, we extract from the results of Sec.~\ref{SecII} the QNM counterpart of the waveforms corresponding to the electromagnetic ringing of the BH. We have gathered all our numerical results and their analysis in Sec.~\ref{SecIV} where we display the regularized multipolar waveforms emitted and compare them with the unregularized counterparts constructed solely from the QNM spectrum. It should be noted that our results are obtained for arbitrary directions of observation and, in particular, outside the orbital plane of the plunging particle. In the late phase of the signals, the exact multipolar waveforms and the associated quasinormal ringings are in excellent agreement. Moreover, our results especially emphasize the impact of higher harmonics on the distortion of the waveforms. In the Conclusion, we summarize the main results obtained in this article and we consider its extension to the much more important problem of the construction of the multipolar gravitational waveforms generated by a massive point particle plunging from slightly below the ISCO into a Schwarzschild BH. In an Appendix, we carefully examine the regularization of the partial amplitudes from both the theoretical and numerical point of view. Indeed, the exact waveforms theoretically constructed in Sec.~\ref{SecII} are integrals over the radial Schwarzschild coordinate which are strongly divergent near the ISCO. For odd perturbations, they can be ``numerically regularized'' by using Levin's algorithm \cite{Levin1996} but, for even perturbations, it is necessary to reduce the degree of divergence of these integrals by successive integrations by parts before applying the same numerical algorithm.

Throughout this article, we adopt units such that $G = c = 1$ and we use the geometrical conventions of Ref.~\cite{Misner:1974qy}.

\section{Electromagnetic field generated by the plunging charged particle}
\label{SecII}

In this section, we shall obtain theoretically the expression of the even- and odd-parity waveforms emitted by a charged point particle plunging from slightly below the ISCO into the BH by working in the frequency domain and using the standard Green's function techniques. Moreover, we shall fix the notations used throughout the whole article.

We first recall that the exterior of the Schwarzschild BH of
mass $M$ is defined by the metric
\begin{equation}\label{Metric_Schwarzschild}
ds^2= -f(r) dt^2+ f(r)^{-1}dr^2+ r^2 d\sigma_2^2
\end{equation}
where $f(r)=(1-2M/r)$ and $d\sigma_2^2=d\theta^2 + \sin^2 \theta d\varphi^2$ denotes the
metric on the unit $2$-sphere $S^2$ and with the Schwarzschild
coordinates $(t,r,\theta,\varphi)$ which satisfy $t \in ]-\infty,
+\infty[$, $r \in ]2M,+\infty[$, $\theta \in [0,\pi]$ and $\varphi
\in [0,2\pi]$. In the following, we shall also use the so-called tortoise coordinate $r_\ast \in ]-\infty,+\infty[$ defined in term of the radial Schwarzschild coordinate $r$ by $dr/dr_\ast=f(r)$ and given by $r_\ast(r)=r+2M \ln[r/(2M)-1]$. We recall that the function $r_\ast=r_\ast(r)$
provides a bijection from $]2M,+\infty[$ to $]-\infty,+\infty[$.

\subsection{Electromagnetic field generated by a moving charged particle}

Electromagnetic radiation of the Schwarzschild BH excited by a charged particle (we denote by $q$ its electric charge) moving along a world line $\gamma$ described by the equations $x_p^\mu = x_p^\mu (\tau)$ (here $\tau$ is the proper time of the charged particle) is governed by the wave equation
\begin{equation}\label{WEQ_potEM}
 \Box A_\mu -  \nabla_\mu \nabla_\nu A^\nu  = -J_\mu.
\end{equation}
Here, $A_\mu$ is the electromagnetic potential generated by the moving particle and $J_\mu$ is the corresponding current given by
\begin{equation}\label{WEQ_currentEM}
J^\mu (x)= q\int_{\gamma}d\tau\, \frac{dx^\mu_p(\tau)}{d\tau} \frac{\delta^{4}(x-x_p(\tau))}{\sqrt{-g(x)}}.
\end{equation}
In order to solve the wave equation (\ref{WEQ_potEM}) in the Schwarzschild spacetime (see, e.g., Ref.~\cite{Ruffini:1972pw}), we can expand the electromagnetic potential $A_\mu$ and the current $J_\mu$ in vector spherical harmonics in the form $A_\mu = A_\mu^{(e)} + A_\mu^{(o)}$ and $J_\mu = J_\mu^{(e)} + J_\mu^{(o)}$. Here, and in the following, the symbols $(e)$ and $(o)$ are respectively associated with even (polar) and odd (axial) objects according they are of even or odd parity in the antipodal transformation on the unit $2$-sphere $S^2$. We have
\begin{subequations}\label{Pot_A}
\begin{equation}\label{Pot_A_even}
A_\mu^{(e)} = \sum\limits_{\ell=0}^{+\infty}\sum\limits_{m=-\ell}^{+\ell}\left(M_t^{\ell m} Y^{\ell m}, M_r^{\ell m} Y^{\ell m}, M^{\ell m} Y_\theta^{\ell m}, M^{\ell m} Y_\varphi^{\ell m}\right)
\end{equation}
and
\begin{equation}\label{Pot_A_odd}
A_\mu^{(o)} = \sum\limits_{\ell=1}^{+\infty}\sum\limits_{m=-\ell}^{+\ell}\left(0, 0, N^{\ell m} X_\theta^{\ell m}, N^{\ell m} X_\varphi^{\ell m}\right)
\end{equation}
\end{subequations}
for the potential and
\begin{subequations}\label{Current_J}
\begin{equation}\label{Current_J_even}
J_\mu^{(e)} = \sum\limits_{\ell=0}^{+\infty}\sum\limits_{m=-\ell}^{+\ell}\left(J_t^{\ell m} Y^{\ell m}, J_r^{\ell m} Y^{\ell m}, J^{\ell m} Y_\theta^{\ell m}, J^{\ell m} Y_\varphi^{\ell m}\right)
\end{equation}
and
\begin{equation}\label{Current_J_odd}
J_\mu^{(o)} = \sum\limits_{\ell=1}^{+\infty}\sum\limits_{m=-\ell}^{+\ell}\left(0, 0, K^{\ell m} X_\theta^{\ell m}, K^{\ell m} X_\varphi^{\ell m}\right)
\end{equation}
for the current.
\end{subequations}
Here, the components $M_t^{\ell m}$, $M_r^{\ell m}$, $M^{\ell m}$ and $N^{\ell m}$ of the potential as well as the components $J_t^{\ell m}$, $J_r^{\ell m}$, $J^{\ell m}$ and $K^{\ell m}$ of the current are functions of $t$ and $r$. The angular functions $Y^{\ell m}(\theta,\varphi)$ are the standard scalar spherical harmonics while the angular functions $Y_\theta^{\ell m}(\theta,\varphi)$, $Y_\varphi^{\ell m}(\theta,\varphi)$, $X_\theta^{\ell m}(\theta,\varphi)$ and $X_\varphi^{\ell m}(\theta,\varphi)$ are the vector spherical harmonics which are given by
\begin{subequations}\label{HSV_even_odd}
\begin{equation}\label{HSV_even}
Y_\theta^{\ell m} = \frac{\partial}{\partial \theta} Y^{\ell m} \quad \text{and} \quad Y_\varphi^{\ell m}= \frac{\partial}{\partial \varphi} Y^{\ell m}
 \end{equation}
for the even vector spherical harmonics and by
\begin{equation}\label{HSV_odd}
X_\theta^{\ell m} = \frac{1}{\sin \theta} \frac{\partial}{\partial \varphi} Y^{\ell m} \quad \text{and} \quad X_\varphi^{\ell m}= - \sin \theta \frac{\partial}{\partial \theta} Y^{\ell m}
 \end{equation}
 \end{subequations}
for the odd ones. It is important to recall that, due to the relation $Y^{\ell -m}=(-1)^m [Y^{\ell m}]^\ast$, the vector spherical harmonics satisfy
\begin{subequations}
\begin{equation}\label{HSV_even_mm}
Y_\theta^{\ell -m} = (-1)^m [Y_\theta^{\ell m}]^\ast  \,\,\, \text{and}  \,\,\, Y_\varphi^{\ell -m} = (-1)^m [Y_\varphi^{\ell m}]^\ast
 \end{equation}
as well as
\begin{equation}\label{HSV_odd_mm}
X_\theta^{\ell -m} = (-1)^m [X_\theta^{\ell m}]^\ast  \,\,\, \text{and}  \,\,\,  X_\varphi^{\ell -m} = (-1)^m [X_\varphi^{\ell m}]^\ast.
 \end{equation}
\end{subequations}

By inserting (\ref{Pot_A}) and (\ref{Current_J}) into the wave equation (\ref{WEQ_potEM}) and using the properties of the spherical harmonics (see, e.g., Refs.~\cite{Martel:2005ir} or \cite{Nagar:2005ea}), we can construct two gauge-invariant functions denoted by $\psi^{(e)}_{\ell m} (t,r)$ and $\psi^{(o)}_{\ell m} (t,r)$ and given by \cite{Ruffini:1972pw}
\begin{equation}\label{Psi_even}
\psi^{(e)}_{\ell m} = r^2 \left(\frac{\partial}{\partial t} M_r^{\ell m} -\frac{\partial}{\partial r} M_t^{\ell m}\right)
 \end{equation}
and
 \begin{equation}\label{Psi_odd}
\psi^{(o)}_{\ell m} = \ell (\ell+1) N^{\ell m}.
 \end{equation}
Here $\ell = 1, 2, 3\dots$ and $m=-\ell, -\ell+1,\dots, +\ell$ and the $(\ell=0,m=0)$ mode has been eliminated by a suitable gauge transformation. These gauge invariant functions satisfy the Regge-Wheeler equation
\begin{subequations} \label{RWem EQ}
\begin{equation} \label{RWem EQa}
\left[- \frac{\partial^2}{\partial t^2} + \frac{\partial^2}{\partial
r_\ast^2} - V_\ell (r)  \right] \psi^{(e/o)}_{\ell m} (t,r) = S^{(e/o)}_{\ell m} (t,r)
\end{equation}
with
\begin{equation} \label{RWem EQb}
V_\ell(r)=f(r) \left(\frac{\ell(\ell+1)}{r^2} \right).
\end{equation}
It is important to note that the potential $V_\ell(r)$ is the same in the two parity sectors.
\end{subequations}
Here the functions $S^{(e)}_{\ell m} (t,r)$ and $S^{(o)}_{\ell m} (t,r)$ are source terms which depend on the trajectory $x_p^\mu = x_p^\mu (\tau)$ of the charged particle and which can be expressed from the components of the current (\ref{Current_J}). They are given by
\begin{equation}\label{Comp_J_Source_even}
S^{(e)}_{\ell m} = -f(r) \left[r^2 \frac{\partial}{\partial t}J_r^{\ell m}- \frac{\partial}{\partial r}\left(r^2 J_t^{\ell m}\right)\right]
\end{equation}
and
\begin{equation}\label{Comp_J_Source_odd}
S^{(o)}_{\ell m} = -\ell (\ell+1) f(r)K^{\ell m}.
\end{equation}

We must point out that the partial amplitudes $\psi^{(e/o)}_{\ell m} (t,r)$ permit us to obtain the electromagnetic field $({\bf E}, {\bf B})$ observed at spatial infinity (i.e., for $r \to +\infty$). We have, in particular, in the usual orthonormalized basis $({\bf {\hat{e}}_r},{\bf {\hat{e}}_\theta},{\bf {\hat{e}}_\varphi})$ of the spherical coordinate system, the electric field which is given in the even sector by
\begin{equation}\label{ChampE_even}
{\mathbf E^{(e)}}=\left| \begin{array}{l}
 E_r^{(e)} = 0 \\
 E_\theta^{(e)} = -\frac{1}{r} \sum\limits_{\ell m}\frac{1}{\ell(\ell+1)}\partial_r\psi^{(e)}_{\ell m} \, Y_\theta^{\ell m} \\
 E_\varphi^{(e)} = -\frac{1}{r \sin\theta} \sum\limits_{\ell m}\frac{1}{\ell(\ell+1)}\partial_r \psi^{(e)}_{\ell m} \,Y_\varphi^{\ell m}, \\
\end{array}
\right.
\end{equation}
and, in the odd sector, by
\begin{equation}\label{ChampE_odd}
{\mathbf E^{(o)}}=\left| \begin{array}{l}
 E_r^{(o)} = 0 \\
 E_\theta^{(o)} = -\frac{1}{r} \sum\limits_{\ell m}\frac{1}{\ell(\ell+1)}\partial_t\psi^{(o)}_{\ell m} \, X_\theta^{\ell m} \\
 E_\varphi^{(o)} = -\frac{1}{r \sin\theta} \sum\limits_{\ell m}\frac{1}{\ell(\ell+1)}\partial_t \psi^{(o)}_{\ell m} \, X_\varphi^{\ell m}. \\
\end{array}
\right.
\end{equation}
In the following, we shall not consider the magnetic field ${\mathbf B}$ because its components, which can be obtained from the Maxwell-Faraday equation, can be expressed in terms of those of the electric field. Indeed, it should be noted that, for $r \to +\infty$, we have $\partial_t\psi^{(e/o)}_{\ell m} = - \partial_r\psi^{(e/o)}_{\ell m}$ and, as a consequence of (\ref{HSV_even}) and (\ref{HSV_odd}), we can write the relations
\begin{equation}\label{Relations_ChampB_even_odd}
B_\theta^{(e/o)} = - E_\varphi^{(e/o)} \quad \mathrm{and} \quad B_\varphi^{(e/o)} = + E_\theta^{(e/o)}.
\end{equation}

In order to solve the Regge-Wheeler equation (\ref{RWem EQ}), we shall work in the frequency domain by writing
\begin{equation}\label{TF_psi}
\psi^{(e/o)}_{\ell m} (t,r) = \frac{1}{\sqrt{2\pi}} \int_{-\infty}^{+\infty} d\omega \, \psi^{(e/o)}_{\omega \ell m} (r) e^{-i\omega t} \end{equation}
and
\begin{equation}\label{TF_sources}
S^{(e/o)}_{\ell m} (t,r) = \frac{1}{\sqrt{2\pi}} \int_{-\infty}^{+\infty} d\omega \, S^{(e/o)}_{\omega \ell m} (r) e^{-i\omega t}. \end{equation}
Then, the Regge-Wheeler equation (\ref{RWem EQ}) reduces to
\begin{equation} \label{RWem EQ_Fourier}
\left[\frac{d^2}{d
r_\ast^2} + \omega^2 - V_\ell (r)  \right] \psi^{(e/o)}_{\omega \ell m} (r) = S^{(e/o)}_{\omega \ell m} (r).
\end{equation}

\subsection{Sources due to the plunging charged particle}

\begin{figure}[h!]
\centering
\includegraphics[scale=0.6]{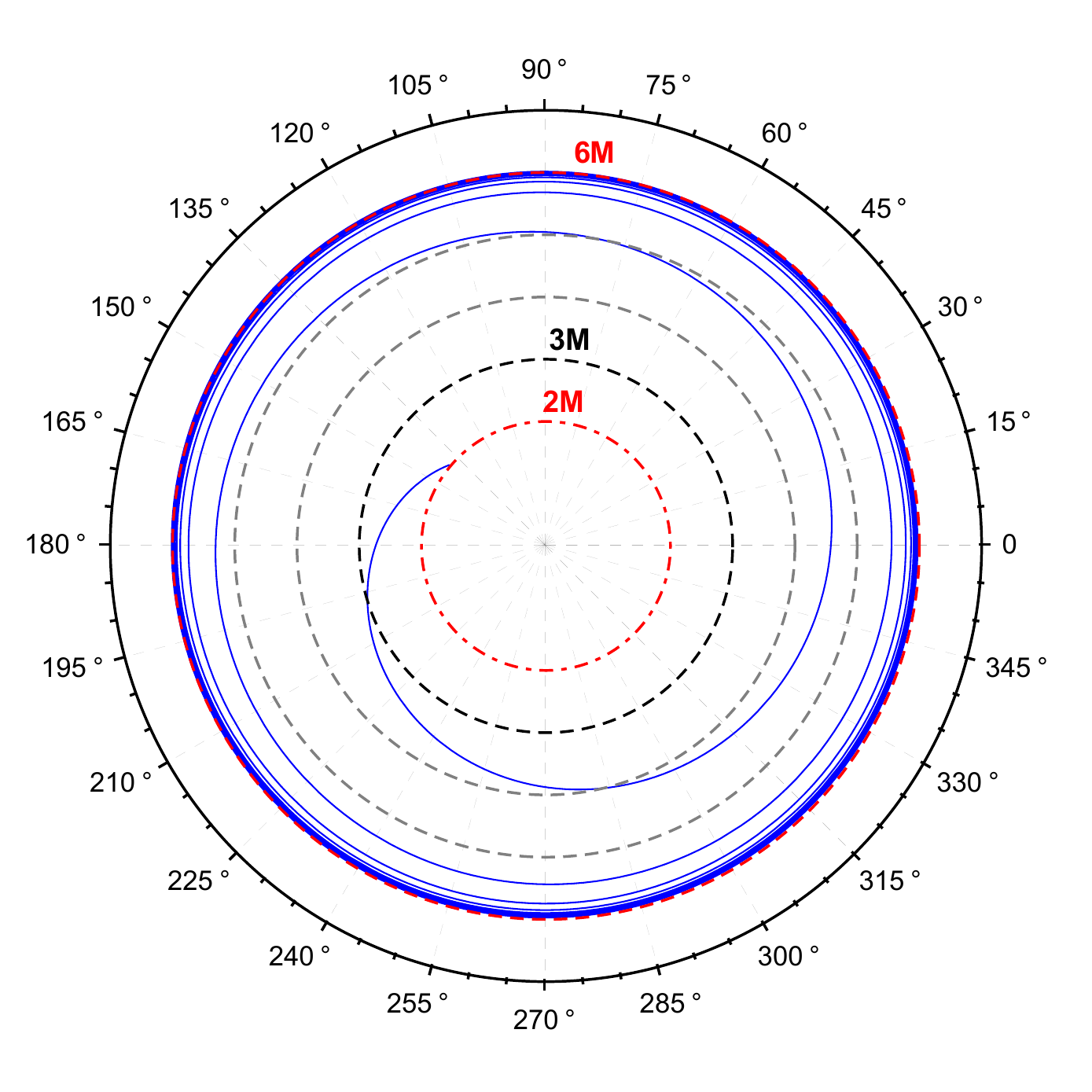}
\setlength\abovecaptionskip{-0.5ex}
\caption{\label{Trajectory_Plung} The plunge trajectory obtained from Eq.~(\ref{trajectory_plung_phi_bis}). Here, we assume that the particle starts at $r=r_\text{\tiny{ISCO}}(1-\epsilon)$ with $\epsilon=10^{-3}$ and we take $\varphi_{0}=0$. The red dashed line at $r=6M$ and the red dot-dashed line at $r=2M$ represent the ISCO and the horizon, respectively, while the black dashed line corresponds to the photon sphere at $r=3M$.}
\end{figure}

In this article, we shall focus on the electromagnetic radiation generated by a particle plunging into the BH from the ISCO at $r_{\mathrm{ISCO}}= 6M$ or, more precisely, from slightly below the ISCO. We denote by $t_{p}(\tau)$, $r_{p}(\tau)$, $\theta_{p}(\tau)$ and $\varphi_{p}(\tau)$ the coordinates of the timelike geodesic followed by the charged particle and, without loss of generality, we consider that its trajectory lies in the BH equatorial plane, i.e., we assume that $\theta_{p}(\tau)=\pi/2$. The geodesic equations are given by \cite{Chandrasekhar:1985kt}
\begin{subequations}
\label{geodesic_equations}
\begin{eqnarray}
& & f(r_{p})\frac{dt_{p}}{d\tau}=\widetilde{E}, \label{geodesic_1} \\
& & r_{p}^{2}\,\,\frac{d\varphi_{p}}{d\tau}=\widetilde{L} \label{geodesic_2}
\end{eqnarray}
\noindent and
\begin{equation}
\label{geodesic_3}
\left(\frac{dr_{p}}{d\tau}\right)^{2} +\frac{\widetilde{L}^{2}}{r_{p}^{2}}f(r_{p})-\frac{2M}{r_{p}}
=\widetilde{E}^{2}-1.
\end{equation}
\end{subequations}
\noindent Here $\widetilde{E}$ and $\widetilde{L}$ are, respectively, the energy and angular momentum per unit mass of the charged particle. These two quantities are conserved along the geodesic and given on the ISCO, i.e., at $r=r_\text{\tiny{ISCO}}$ with
\begin{equation} \label{r_isco}
r_\text{\tiny{ISCO}}=6M,
\end{equation}
by
\begin{equation} \label{EetL_isco}
\widetilde{E}=\frac{2\sqrt{2}}{3} \quad \mathrm{and} \quad \widetilde{L}=2\sqrt{3}M.
\end{equation}
By substituting (\ref{EetL_isco}) into the geodesic equations (\ref{geodesic_1})-(\ref{geodesic_3}), we obtain after integration
\begin{eqnarray}
\label{trajectory_plung}
\frac{t_{p}(r)}{2M}\!&=&\!\frac{2\sqrt{2}\left(r-24M\right)}{2M\left(6M/r-1\right)^{1/2}}-22\sqrt{2}
\tan^{-1}\!\!\left[\left(6M/r-1\right)^{1/2}\right]\nonumber\\
&&+ 2\tanh^{-1}\!\!\left[\frac{1}{\sqrt{2}}\left(6M/r-1\right)^{1/2}\right]+  \frac{t_{0}}{2M}
\end{eqnarray}
\noindent and
\begin{equation}
\label{trajectory_plung_phi}
\varphi_{p}(r)=-\frac{2\sqrt{3}}{\left(6M/r-1\right)^{1/2}}+\varphi_{0}
\end{equation}
\noindent where $t_{0}$ and $\varphi_{0}$ are two arbitrary integration constants. From (\ref{trajectory_plung_phi}), we can write the spatial trajectory of the plunging particle in the form
\begin{equation}
\label{trajectory_plung_phi_bis}
r_{p}(\varphi)=\frac{6M}{[1+12/(\varphi-\varphi_{0})^{2}]}.
\end{equation}
This trajectory is displayed in Fig.~\ref{Trajectory_Plung}.

For the plunging charged particle, the sources $S^{(e/o)}_{\ell m} (t,r)$ can be obtained from the expansions (\ref{Current_J}) by inserting the geodesic equations (\ref{geodesic_1})-(\ref{geodesic_3}) into (\ref{WEQ_currentEM}) and by using (\ref{EetL_isco}) as well as the change of variable $\tau \to r_p(\tau)$. From (\ref{WEQ_currentEM}) we have
\begin{eqnarray}\label{WEQ_currentEM_plunging}
& & J^\mu (x)= q\frac{dx^\mu_p}{d\tau}(r)  \left[\frac{ dr_p}{d\tau} (r)\right]^{-1}\nonumber \\
& & \qquad \qquad \times \frac{\delta[t-t_p(r)] \delta[\theta- \pi/2] \delta[\varphi-\varphi_p(r)]}{r^2 \sin \theta}
\end{eqnarray}
which permits us to write
\begin{widetext}
\begin{eqnarray} \label{Source_even}
 & & S^{(e)}_{\ell m} (t,r) = q \left[Y^{\ell m}(\pi/2,0)\right]^* f(r) \left[ \left(\frac{18\sqrt{2}M}{r^2(6M/r-1)^{5/2}}+i m \frac{12\sqrt{6}M}{r^2(6M/r-1)^{3}}\right)\delta\left[t-t_{p}(r)\right] \right. \nonumber \\
 & & \left. \phantom{\frac{18\sqrt{2}M}{r^2(6M/r-1)^{5/2}}} \qquad\qquad   +\frac{9(r^2+12M^2)}{r^2(6M/r-1)^{3}}\delta'\left[t - t_{p}(r)\right]\right]e^{- i m \varphi_{p}(r)}
 \end{eqnarray}
and
\begin{equation}
 \label{Source_odd}
 S^{(o)}_{\ell m} (t,r) = q \left[X_\varphi^{\ell m}(\pi/2,0)\right]^* f(r)\frac{6\sqrt{3}M}{r^2(6M/r-1)^{3/2}} \delta\left[t-t_{p}(r)\right] e^{- i m \varphi_{p}(r)}
 \end{equation}
and we have for their Fourier components defined by (\ref{TF_sources})
\begin{equation}
 \label{SourceTR_even}
 S^{(e)}_{\omega \ell m} (r) = \frac{q}{\sqrt{2\pi}} \left[Y^{\ell m}(\pi/2,0)\right]^* f(r) \left[ \frac{18\sqrt{2}M}{r^2(6M/r-1)^{5/2}}+i m \frac{12\sqrt{6}M}{r^2(6M/r-1)^{3}}  -i\omega \frac{9(r^2+12M^2)}{r^2(6M/r-1)^{3}}\right]e^{i[\omega t_p(r) - m \varphi_{p}(r)]}
 \end{equation}
and
\begin{equation}
 \label{SourceTR_odd}
 S^{(o)}_{\omega \ell m} (r) = \frac{q}{\sqrt{2\pi}} \left[X_\varphi^{\ell m}(\pi/2,0)\right]^* f(r) \left[\frac{6\sqrt{3}M}{r^2(6M/r-1)^{3/2}} \right]e^{i[\omega t_p(r) - m \varphi_{p}(r)]}.
 \end{equation}
\end{widetext}

It is interesting to remark that, as a consequence of (\ref{HSV_even_mm}) and (\ref{HSV_odd_mm}), we have
\begin{equation}\label{S_omega_mm}
 S^{(e/o)}_{\omega \ell -m} = (-1)^m [ S^{(e/o)}_{-\omega \ell m}]^\ast.
\end{equation}
Furthermore, it is important to note that the coefficient $Y^{\ell m}(\pi/2,0)$ appearing in Eqs.~(\ref{Source_even}) and (\ref{SourceTR_even}) and the coefficient $X_\varphi^{\ell m}(\pi/2,0)$ appearing in Eqs.~(\ref{Source_odd}) and (\ref{SourceTR_odd}) are given by
\begin{eqnarray} \label{PourSe_Y}
& & Y^{\ell m}(\pi/2,0)=\frac{2^m}{\sqrt{\pi}}\sqrt{\frac{2\ell +1}{4\pi}\frac{(\ell-m)!}{(\ell+m)!}} \nonumber \\
& & \qquad \times \frac{\Gamma[\ell/2 + m/2 +1/2]}{\Gamma[\ell/2 - m/2 +1]} \cos\left[(\ell+m)\pi/2\right]
\end{eqnarray}
and
\begin{eqnarray} \label{PourSo_X}
& & X_\varphi^{\ell m}(\pi/2,0)=\frac{2^{m+1}}{\sqrt{\pi}}\sqrt{\frac{2\ell +1}{4\pi}\frac{(\ell-m)!}{(\ell+m)!}} \nonumber \\
& & \qquad \times \frac{\Gamma[\ell/2 + m/2 +1]}{\Gamma[\ell/2 - m/2 +1/2]} \sin\left[(\ell+m)\pi/2\right].
\end{eqnarray}
As a consequence, $Y^{\ell m}(\pi/2,0)$ and hence the sources (\ref{Source_even}) and (\ref{SourceTR_even}) vanish for $\ell+m$ odd while $X_\varphi^{\ell m}(\pi/2,0)$ and hence the sources (\ref{Source_odd}) and (\ref{SourceTR_odd}) vanish for $\ell+m$ even. Due to these results,  we have to only consider the couples $(\ell,m)$ with $\ell+m$ even in the expression (\ref{ChampE_even}) of the electric field ${\mathbf E^{(e)}}$ and the couples $(\ell,m)$ with $\ell+m$ odd in the expression (\ref{ChampE_odd}) of the electric field ${\mathbf E^{(o)}}$.

\subsection{Construction of the partial amplitudes $\psi^{(e/o)}_{\ell m} (t,r)$}

The Regge-Wheeler equation (\ref{RWem EQ_Fourier}) can be solved by using the machinery of Green's functions (see Ref.~\cite{MorseFeshbach1953} for generalities on this topic and, e.g., Ref.~\cite{Breuer:1974uc} for its use in the context of BH physics). We consider the Green's function $G_{\omega\ell}(r_{*},{r}_{*}')$ defined by
\begin{equation}
\label{Green_Function_1}
\left[\frac{d^{2}}{dr_{\ast}^{2}}+\omega^{2}-V_{\ell}(r)\right]G_{\omega\ell}(r_{*},{r}_{*}')=-\delta(r_{*}-{r}_{*}')
\end{equation}
which can be written as
\begin{equation}
\label{Green_Function_2}
G_{\omega\ell}(r_{*},{r}_{*}')=-\frac{1}{W_{\ell}(\omega)}
\left\{
\begin{aligned}
&\!\!\phi_{\omega\ell}^{\mathrm{in}}(r_{*})\,\phi_{\omega\ell}^{\mathrm{up}}({r}_{*}'),\!\!&r_{*}<{r}_{\ast}',\\
&\!\!\phi_{\omega\ell}^{\mathrm{up}}(r_{*})\,\phi_{\omega\ell}^{\mathrm{in}}({r}_{*}'),\!\!&r_{*}>{r}_{\ast}'.
\end{aligned}
\right.
\end{equation}
\noindent Here $W_{\ell}(\omega)$ denotes the Wronskian of the functions $\phi_{\omega\ell}^{\mathrm {in}}$ and $\phi_{\omega\ell}^{\mathrm {up}}$. These two functions are linearly independent solutions of the homogenous Regge-Wheeler equation
\begin{equation}
\label{H_RW_equation}
\left[\frac{d^{2}}{dr_{\ast}^{2}}+\omega^{2}-V_{\ell}(r)\right]\phi_{\omega\ell}= 0.
\end{equation}
$\phi_{\omega\ell}^{\mathrm {in}}$ is defined by its purely ingoing behavior at the event horizon $r=2M$ (i.e., for $r_\ast \to -\infty$)
\begin{subequations}
\label{bc_in}
\begin{equation}\label{bc_1_in}
\phi^\mathrm{in}_{\omega \ell} (r)\scriptstyle{\underset{r_\ast \to -\infty}{\sim}} \displaystyle{e^{-i\omega r_\ast}}
\end{equation}
while, at spatial infinity $r \to +\infty$ (i.e., for $r_\ast \to +\infty$), it has an
asymptotic behavior of the form
\begin{equation}\label{bc_2_in}
\phi^\mathrm{in}_{\omega  \ell}(r) \scriptstyle{\underset{r_\ast \to +\infty}{\sim}}
\displaystyle{ A^{(-)}_\ell (\omega) e^{-i\omega r_\ast} + A^{(+)}_\ell (\omega) e^{+i\omega r_\ast}}.
\end{equation}
\end{subequations}
Similarly, $\phi^\mathrm{up}_{\omega \ell }$ is defined by its purely outgoing behavior at spatial infinity
\begin{subequations}
\label{bc_up}
\begin{equation}\label{bc_1_up}
\phi^\mathrm{up}_{\omega \ell} (r)\scriptstyle{\underset{r_\ast \to +\infty}{\sim}}
 \displaystyle{ e^{+i\omega r_\ast}}
\end{equation}
and, at the horizon, it has an asymptotic behavior of the form
\begin{equation}\label{bc_2_up}
\phi^\mathrm{up}_{\omega \ell }(r) \scriptstyle{\underset{r_\ast \to -\infty}{\sim}}\displaystyle{
B^{(-)}_\ell (\omega) e^{-i\omega r_\ast}  + B^{(+)}_\ell (\omega) e^{+i\omega r_\ast}}.
\end{equation}
\end{subequations}
In the previous expressions, the coefficients $A^{(-)}_\ell (\omega)$, $A^{(+)}_\ell (\omega)$, $B^{(-)}_\ell (\omega)$ and $B^{(+)}_\ell (\omega)$ are complex amplitudes. By evaluating the Wronskian $W_\ell (\omega)$ at $r_\ast \to -\infty$ and $r_\ast \to +\infty$, we obtain
\begin{equation}
\label{Well}
W_\ell (\omega) =2i\omega A^{(-)}_\ell (\omega) = 2i\omega B^{(+)}_\ell (\omega).
\end{equation}
Here, it is worth noting some important properties of the coefficients $A^{(-)}_\ell (\omega)$ and $A^{(+)}_\ell (\omega)$ and of the function $\phi^\mathrm{in}_{\omega  \ell}(r)$ that we will use extensively later. They are a direct consequence of Eqs.~(\ref{H_RW_equation}) and (\ref{bc_in}) and they are valid whether $\omega$ is real or complex:
\begin{equation}\label{PhiIN_mm}
\phi_{-\omega \ell}^{\mathrm {in}}(r) = [\phi_{\omega\ell}^{\mathrm {in}}(r)]^\ast \quad \mathrm{and} \quad A_{\ell}^{(\pm)}(-\omega)=[A_{\ell}^{(\pm)}(\omega)]^\ast.
\end{equation}

Now, by using the Green's function (\ref{Green_Function_2}), we can show that the solution of the Regge-Wheeler equation with source (\ref{RWem EQ_Fourier}) is given by
\begin{subequations}\label{G_Sol_RW_Eq_en_ast}
\begin{eqnarray}
& & \psi^{(e/o)}_{\omega\ell m}(r)=- \int_{-\infty}^{+\infty}d{r}_{*}'\,G_{\omega\ell}(r_{*},{r}_{*}')
S^{(e/o)}_{\omega\ell m}({r}_{*}') \label{G_Sol_RW_Eq_en_ast_a}\\
& & \phantom{\psi^{(e/o)}_{\omega\ell m}(r)} = - \int_{2M}^{6M} \frac{dr'}{f(r')} \,  G_{\omega\ell}(r,r')
S^{(e/o)}_{\omega\ell m}(r'). \label{G_Sol_RW_Eq_en_ast_b}
\end{eqnarray}
\end{subequations}
For $r \to +\infty$, the solution (\ref{G_Sol_RW_Eq_en_ast_b}) reduces to the asymptotic expression
\begin{eqnarray}
\label{Partial_Response_1}
& & \psi^{(e/o)}_{\omega\ell m}(r)= \frac{e^{+i \omega r_\ast }}{2i\omega A^{(-)}_\ell (\omega)}  \int_{2M}^{6M} \frac{dr'}{f(r')} \,\phi_{\omega\ell}^{\mathrm{in}}(r')
\,S^{(e/o)}_{\omega\ell m}(r'). \nonumber \\
& &
\end{eqnarray}
This result is a consequence of Eqs.~(\ref{Green_Function_2}), (\ref{bc_1_up}) and (\ref{Well}).

We can now obtain the solution of the Regge-Wheeler equation (\ref{RWem EQ} by inserting (\ref{Partial_Response_1}) into (\ref{TF_psi}) and we have for the $(\ell,m)$ waveform in the time domain
\begin{eqnarray}
\label{partial_response_def}
& & \psi^{(e/o)}_{\ell m}(t,r) =  \frac{1}{\sqrt{2\pi}} \int_{-\infty}^{+\infty} d\omega  \left(\frac{e^{- i \omega [t-r_\ast (r)]}}{2 i \omega A_{\ell}^{(-)}(\omega)}\right)\nonumber\\
& & \qquad\qquad \qquad \quad \times\,  \int_{2M}^{6M} \frac{dr'}{f(r')} \,\phi_{\omega\ell}^{\mathrm{in}}(r')
\,S^{(e/o)}_{\omega\ell m}(r').
\end{eqnarray}

Here it is important to note that these partial waveforms satisfy
\begin{equation}\label{Psi_t_mm}
\psi^{(e/o)}_{\ell -m} = (-1)^m [\psi^{(e/o)}_{\ell m}]^\ast.
\end{equation}
This is a direct consequence of the definition (\ref{TF_psi}) and of the relation
\begin{equation}\label{Psi_om_mm}
\psi^{(e/o)}_{\omega \ell -m} = (-1)^m [\psi^{(e/o)}_{\omega \ell m}]^\ast
\end{equation}
which is easily obtained from (\ref{Partial_Response_1}) and (\ref{S_omega_mm}) by noting that the solution $\phi_{\omega\ell}^{\mathrm {in}}$ of the problem (\ref{H_RW_equation})-(\ref{bc_in}) and the associated coefficient $A_{\ell}^{(-)}(\omega)$ satisfy (\ref{PhiIN_mm}). The relations (\ref{Psi_t_mm}), (\ref{HSV_even_mm}) and (\ref{HSV_odd_mm}) permit us to check that the electric fields (\ref{ChampE_even}) and (\ref{ChampE_odd}) are purely real.

\section{Quasinormal ringings due to the plunging charged particle}
\label{SecIII}

\begingroup
\begin{table}[htp]
\caption{\label{tab:table1} The first quasinormal frequencies $\omega_{\ell n}$ and the associated excitation factors $\mathcal{B}_{\ell n}$.}
\smallskip
\centering
\begin{ruledtabular}
\begin{tabular}{ccc}
 $(\ell, n)$ & $2M \omega_{\ell n}$ & $\mathcal{B}_{\ell n}$
 \\ \hline
 $(1,1)$  & $0.496527 - 0.184975 i$  & $-0.161402 + 0.011856 i$   \\
$(2,1)$  & $0.915191 - 0.190009 i$  & $\phantom{-}  0.121187 + 0.018638 i$   \\
 $(3,1)$  & $1.313800 - 0.191232 i$  & $-0.093439 - 0.043528 i$   \\
 $(4,1)$  & $1.706190 - 0.191720 i$  & $\phantom{-} 0.067181 + 0.060960 i$   \\
 $(5,1)$  & $2.095830 - 0.191963 i$  & $-0.041097 - 0.070922$   \\
 $(6,1)$  & $2.483992 - 0.192102 i$  & $\phantom{-}0.015879 + 0.073663 i$ \\
$(7,1)$  & $2.871282 - 0.192189 i$  & $\phantom{-}0.007211 - 0.069753 i$\\
 $(8,1)$  & $3.258010 - 0.192247 i$  & $-0.026845 + 0.060132 i$\\
 $(9,1)$  & $3.644350 - 0.192288 i$  & $\phantom{-}0.041902 - 0.046078 i$ \\
$(10,1)$ & $4.030411 - 0.192317 i$  & $-0.051579 + 0.029124 i$ \\
\end{tabular}
\end{ruledtabular}
\end{table}
\endgroup

\begin{table*}[htp]
\caption{\label{tab:table2} The excitation coefficients $\mathcal{C}_{\ell m n}^{(e/o)}$ and  $\mathcal{D}_{\ell m n}^{(e/o)}$ corresponding to the quasinormal frequencies $\omega_{\ell n}$ and the excitation factors $\mathcal{B}_{\ell n}$ of Table \ref{tab:table1}.}
\smallskip
\centering
\resizebox{\textwidth}{!}{%
\begin{tabular}{ccccc}
\hline
\hline
 $(\ell, n)$ &$\mathcal{C}_{\ell \ell n}^{(e)}$ & $\mathcal{D}_{\ell \ell n}^{(e)}$ & $\mathcal{C}_{\ell {\ell-1} n}^{(o)}$ & $\mathcal{D}_{\ell {\ell-1} n}^{(o)}$
 \\ \hline
$(1,1)$   & $\phantom{-} 7.6654\times10^{-7} + 1.5847\times10^{-6} i$   & $-8.0630\times10^{-8} - 6.3603\times10^{-8} i$  & $\phantom{-} 4.2348\times10^{-8} - 3.4302\times10^{-7} i$  & $\phantom{-} 4.2348\times10^{-8} + 3.4302\times10^{-7} i$    \\
 $(2,1)$ & $-3.4976\times10^{-7} - 3.8945\times10^{-8} i$  & $\phantom{-}1.7576\times10^{-9} + 4.8694\times10^{-10} i$  & $\phantom{-} 5.6771\times10^{-8} + 3.2810\times10^{-8} i$  & $-1.5195\times10^{-9} - 6.4336\times10^{-9} i$   \\
 $(3,1)$ & $\phantom{-} 1.3005\times10^{-7} + 1.7161\times10^{-8} i$  & $-1.2021\times10^{-11} + 6.9750\times10^{-11} i$   & $-2.1452\times10^{-8} - 1.0984\times10^{-8} i$ & $\phantom{-} 2.4896\times10^{-10} - 1.8725\times10^{-11} i$  \\
 $(4,1)$   & $-5.4485\times10^{-8} - 2.7938\times10^{-8} i$  & $-1.1847\times10^{-11} - 2.9359\times10^{-11} i$  & $\phantom{-} 7.8556\times10^{-9} + 7.7218\times10^{-9} i$  & $\phantom{-} 4.3971\times10^{-12} + 1.1610\times10^{-11} i$    \\
 $(5,1)$   & $\phantom{-} 1.8429\times10^{-8} + 2.6355\times10^{-8} i$  & $-1.2915\times10^{-11} + 1.4836\times10^{-11} i$  & $-1.8398\times10^{-9} - 5.3328\times10^{-9} i$  & $-3.1048\times10^{-13} + 7.3141\times10^{-13} i$    \\
 $(6,1)$   &$-8.3837\times10^{-10} - 1.8160\times10^{-8} i$   & $
\phantom{-}1.3477\times10^{-11} + 1.1666\times10^{-12} i$   & $-6.5998\times10^{-10} + 3.0322\times10^{-9} i$  &   $-2.9864\times10^{-14} + 1.8910\times10^{-13} i$ \\
 $(7,1)$ & $-5.7761\times10^{-9} + 9.1300\times10^{-9} i$   &$-5.1587\times10^{-12} - 8.3375\times10^{-12} i$   & $\phantom{-} 1.3014\times10^{-9} - 1.2348\times10^{-9} i$ &   $-6.0170\times10^{-14} + 1.2033\times10^{-13} i$\\
 $(8,1)$  & $\phantom{-} 6.2180\times10^{-9} - 2.3858\times10^{-9} i$   & $-2.6190\times10^{-12} + 6.9236\times10^{-12} i$   & $-1.0671\times10^{-9} + 1.3839\times10^{-10} i$ &  $-7.8861\times10^{-14} + 5.7674\times10^{-14} i$  \\
 $(9,1)$ & $-4.0398\times10^{-9} - 1.2242\times10^{-9} i$   & $\phantom{-} 5.5052\times10^{-12} - 1.7189\times10^{-12} i$ & $\phantom{-} 5.7503\times10^{-10} + 3.3213\times10^{-10} i$  & $-7.2057\times10^{-14} + 1.0705\times10^{-14} i$  \\
 $(10,1)$& $\phantom{-} 1.5465\times10^{-9} + 2.2554\times10^{-9} i$  & $-3.8100\times10^{-12} - 2.5910\times10^{-12} i$  & $-1.5478\times10^{-10} - 3.8961\times10^{-10} i$  & $-5.2748\times10^{-14} - 1.8307\times10^{-14} i$  \\
\hline
\hline
\end{tabular}%
}
\end{table*}

In this section, we shall explain how to construct the quasinormal ringings associated with the electric fields (\ref{ChampE_even}) and (\ref{ChampE_odd}). Of course, they can be obtained by summing over the ringings associated with all the partial amplitudes $\psi^{(e/o)}_{\ell m}(t,r)$. In order to extract from these partial amplitudes the corresponding quasinormal ringings $\psi^{\text{\tiny{QNM}} \, (e/o)}_{\ell m}(t,r)$, the contour of integration over $\omega$ in Eq.~(\ref{partial_response_def}) may be ``deformed'' (see, e.g., Ref.~\cite{Leaver:1986gd}). This deformation permits us to capture the zeros of the Wronskian (\ref{Well}) lying in the lower part of the complex $\omega$ plane and which are the complex frequencies $\omega_{\ell n}$ of the $(\ell,n)$ QNMs. We note that, for a given $\ell$, $n=1$ corresponds to the fundamental QNM (i.e., the least damped one) while $n=2, 3, \dots$ to the overtones. We also recall that the spectrum of the quasinormal frequencies is symmetric with respect to the imaginary axis, i.e., that if $\omega_{\ell n}$ is a quasinormal frequency lying in the fourth quadrant, $-\omega_{\ell n}^{*}$ is the symmetric quasinormal frequency lying in the third one. We easily obtain
\begin{equation}
\label{partial_response_QNM_1}
\psi^{\text{\tiny{QNM}} \, (e/o)}_{\ell m}(t,r) = \sum^{+\infty}_{n=1} \psi^{\text{\tiny{QNM}} \, (e/o)}_{\ell m n}(t,r)
\end{equation}
with
\begin{eqnarray}
\label{partial_response_QNM_2}
 \psi^{\text{\tiny{QNM}} \, (e/o)}_{\ell m n}(t,r) = && -\sqrt{2\pi}\left({\cal{C}}^{(e/o)}_{\ell m n}\,\, e^{-i \omega_{\ell n}[t-r_\ast(r)]}\,\,\vphantom{e^{i \omega_{\ell n}^{\ast}t}}\right.\nonumber\\
&&\left.+\,\,{\cal{D}}^{(e/o)}_{\ell m n}e^{+i \omega^\ast_{\ell n}[t-r_\ast(r)]} \right).
\end{eqnarray}
\noindent In this expression, ${\cal{C}}^{(e/o)}_{\ell m n}$ and ${\cal{D}}^{(e/o)}_{\ell m n}$ denote the extrinsic excitation coefficients (see, e.g., Refs.~\cite{Leaver:1986gd,Berti:2006wq,Zhang:2013ksa} for more details on this concept). They are here defined by
\begin{subequations}\label{excitation_coeffs}
\begin{equation}
\label{excitation_coeff_C}
{\cal{C}}^{(e/o)}_{\ell m n}={\cal{B}}_{\ell n}\left[\int_{2M}^{6M} \frac{dr'}{f(r')} \, \frac{\phi_{\omega \ell}^{\mathrm{in}}(r') }{A_{\ell}^{(+)}(\omega)}S^{(e/o)}_{\omega \ell m}(r') \right]_{\omega=\omega_{\ell n}}
\end{equation}
and
\begin{equation}
\label{excitation_coeff_D}
{\cal{D}}^{(e/o)}_{\ell m n}={\cal{B}}_{\ell n}^\ast \left[\int_{2M}^{6M}  \frac{dr'}{f(r')} \, \frac{\phi_{\omega \ell}^{\mathrm{in}}(r') }{A_{\ell}^{(+)}(\omega)}S^{(e/o)}_{\omega \ell m}(r') \right]_{\omega=-\omega_{\ell n}^{*}}
\end{equation}
\end{subequations}
while
\begin{equation}
\label{excitation_factor}
{\cal{B}}_{\ell n}=\left[\frac{1}{2 \omega}\,\,\frac{A_{\ell}^{(+)}(\omega)} {\frac{d}{d \omega} A_{\ell}^{(-)}(\omega)} \right]_{\omega=\omega_{\ell n}}
\end{equation}
is the so-called excitation factor associated with the $(\ell,n)$ QNM of complex frequency $\omega_{\ell n}$. The first term in the right hand side (r.h.s.)~of Eq.~(\ref{partial_response_QNM_2}) is the contribution of the quasinormal frequency $\omega_{\ell n}$ lying in the fourth quadrant of the $\omega$ plane while the second one is the contribution of $-\omega_{\ell n}^{\ast}$, i.e., its symmetric with respect to the imaginary axis. In front of the bracket in the r.h.s.~of Eq.~(\ref{excitation_coeff_D}), the coefficient ${\cal{B}}_{\ell n}^\ast$ is nothing else than the excitation factor associated with the $(\ell,n)$ QNM of complex frequency $-\omega_{\ell n}^{\ast}$. It is obtained from (\ref{excitation_factor}) by using the properties (\ref{PhiIN_mm}). A few remarks are in order:

\begin{enumerate}[label=(\arabic*)]
   \item The excitation coefficients  (\ref{excitation_coeff_C}) and  (\ref{excitation_coeff_D}) depend on the parity sector because they are constructed from the sources. This is not the case for the excitation factor (\ref{excitation_factor}): indeed, it is defined only from the Regge-Wheeler equation (\ref{RWem EQ_Fourier}) without source term  and where the potential $V_\ell(r)$ is identical in the two parity sectors.

   \item  In our problem, the spherical symmetry of the Schwarzschild BH is broken due to the asymmetric plunging trajectory. It is this dissymmetry which, in connection with the presence of the azimuthal number $m$, forbids us to gather the two terms in Eq.~(\ref{partial_response_QNM_2}).

       \item  It is however important to note that the excitation coefficients (\ref{excitation_coeffs}) are related by
\begin{equation}
\label{excitation_coeffs_CetD_prop}
{\cal{C}}^{(e/o)}_{\ell -m n}= (-1)^m \left[ {\cal{D}}^{(e/o)}_{\ell m n} \right]^\ast
\end{equation}
[this is due to the properties (\ref{PhiIN_mm})] and hence that the quasinormal waveforms (\ref{partial_response_QNM_1}) satisfy
\begin{equation}\label{PsiQNM_t_mm}
\psi^{\text{\tiny{QNM}} \, (e/o)}_{\ell -m} = (-1)^m [\psi^{\text{\tiny{QNM}} \, (e/o)}_{\ell m}]^\ast.
\end{equation}
The quasinormal electric fields obtained from (\ref{ChampE_even}) and (\ref{ChampE_odd}) by replacing $\psi^{(e/o)}_{\ell m}(t,r)$ with $\psi^{\text{\tiny{QNM}} \, (e/o)}_{\ell m}(t,r)$ are then purely real as a consequence of the relations (\ref{PsiQNM_t_mm}), (\ref{HSV_even_mm}) and (\ref{HSV_odd_mm}).

   \item The ringings $\psi^{\text{\tiny{QNM}} \, (e/o)}_{\ell m n}(t,r)$ and $\psi^{\text{\tiny{QNM}} \, (e/o)}_{\ell m}(t,r)$ do not provide physically relevant results at ``early times'' due to their  exponentially divergent behavior as $t$ decreases. It is necessary to determine, from physical considerations (see below), the time beyond which these quasinormal waveforms can be used, i.e., the starting time $t_\mathrm{start}$ of the BH ringings.

\end{enumerate}

\begin{figure*}
\centering
 \includegraphics[scale=0.55]{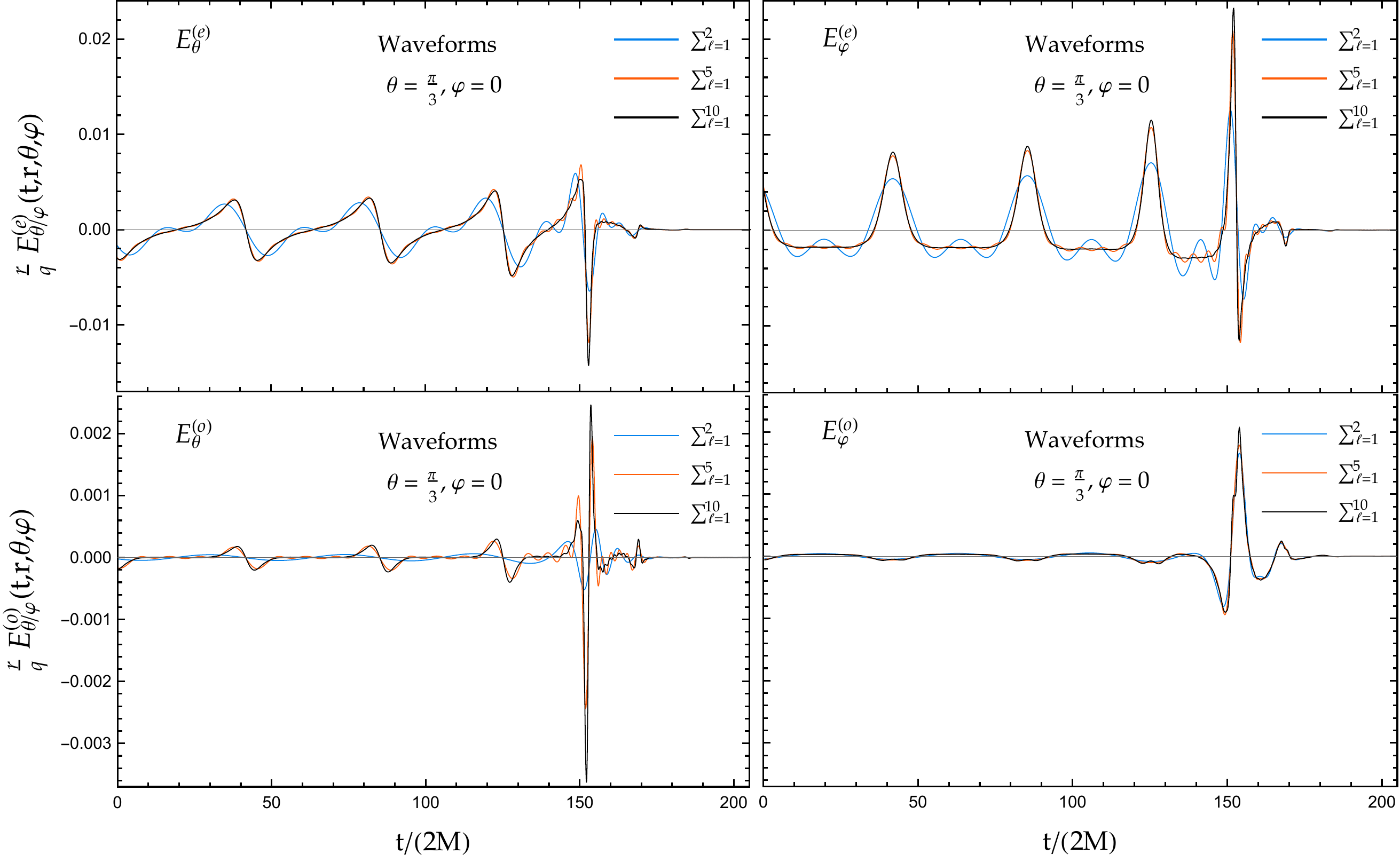}
\caption{\label{E_Sum_ell} Components $E_{\theta/\varphi}^{(e/o)}$ of the electric field observed at infinity in the direction $(\theta=\pi/3,\varphi=0)$ above the orbital plane of the plunging particle. We study the influence of the number of modes on the distortion of the waveforms.}
\end{figure*}

\begin{figure*}
\centering
 \includegraphics[scale=0.5]{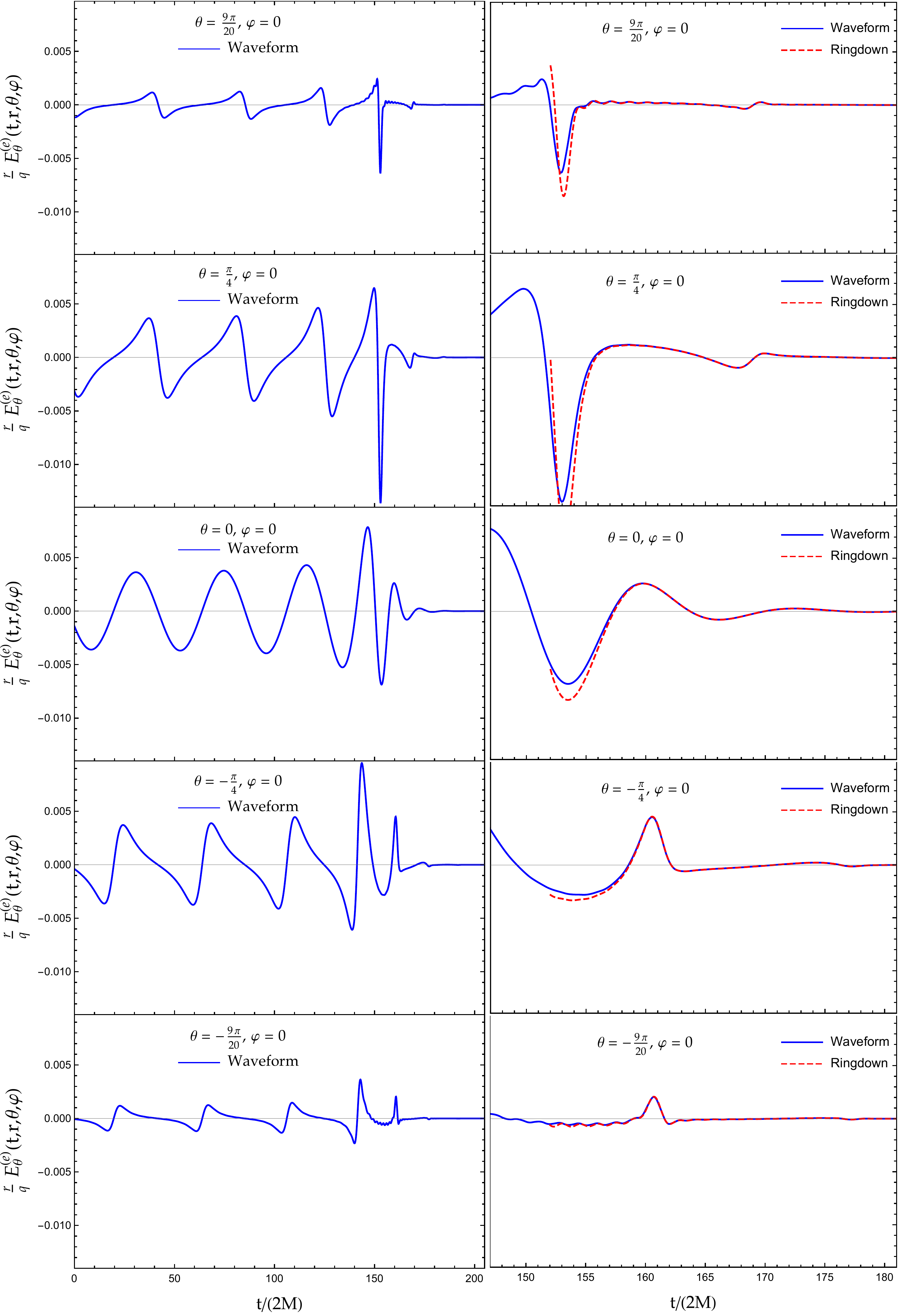}
\caption{\label{Etheta_Waveform_Vs_QNM_even} Electric field $E_\theta^{(e)}$ observed at infinity for various directions above the orbital plane of the plunging particle. We consider $\varphi=0$ and we study the distortion of the multipolar waveform and of the associated quasinormal ringdown when $\theta$ varies between $-\pi/2$  and $+\pi/2$. We note that $E_\theta^{(e)}$ vanishes for $\theta = \pm \pi/2$ and that, for $\theta=0$, only the $(\ell=1,m=\pm 1)$ modes contribute to the signal.}
\end{figure*}

\begin{figure*}
\centering
 \includegraphics[scale=0.5]{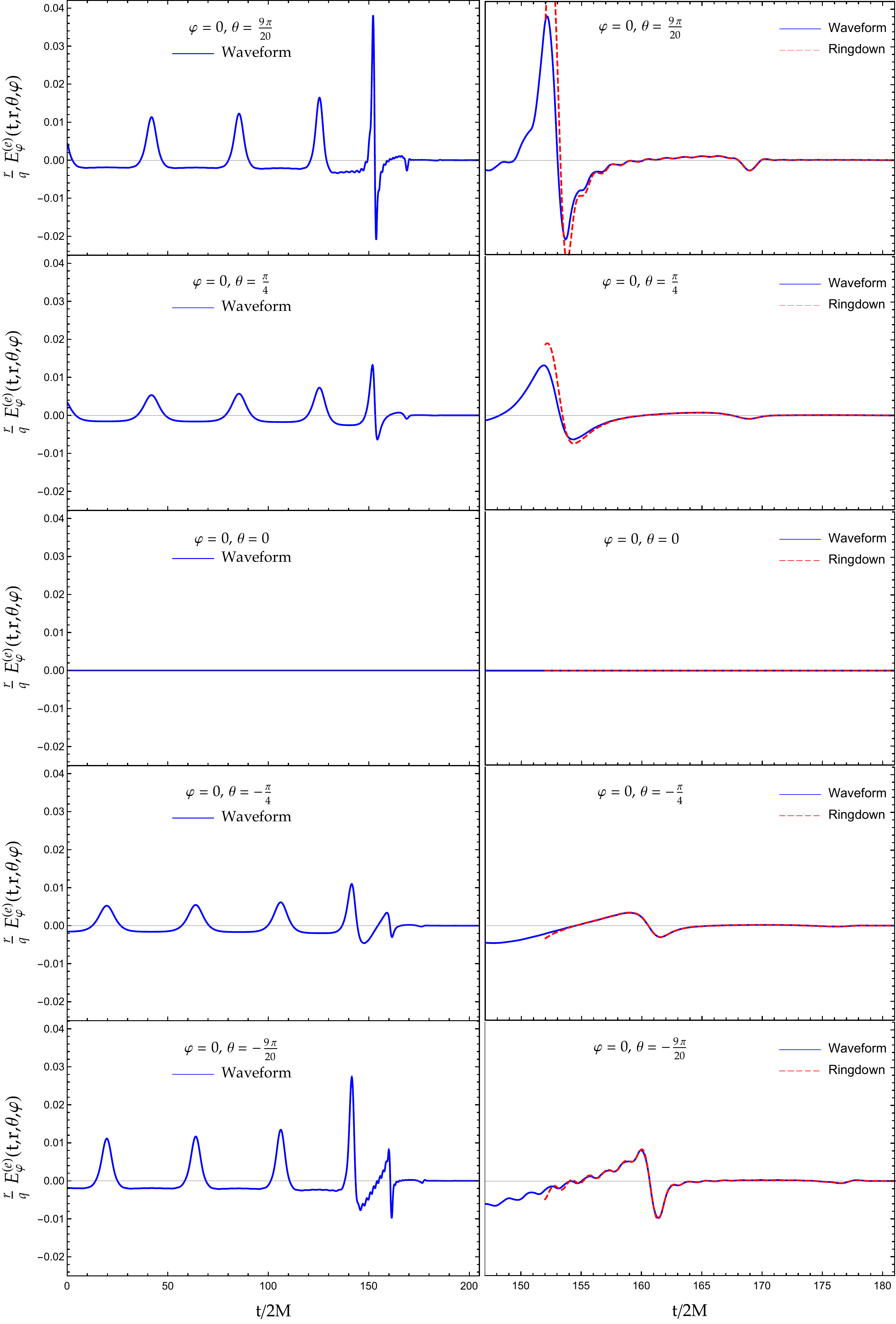}
\caption{\label{Ephi_Waveform_Vs_QNM_even} Electric field $E_\varphi^{(e)}$ observed at infinity for various directions above the orbital plane of the plunging particle. We consider $\varphi=0$ and we study the distortion of the multipolar waveform and of the associated quasinormal ringdown when $\theta$ varies between $-\pi/2$  and $+\pi/2$. We note that $E_\varphi^{(e)}$ vanishes for $\theta = 0$.}
\end{figure*}

\begin{figure*}
\centering
 \includegraphics[scale=0.55]{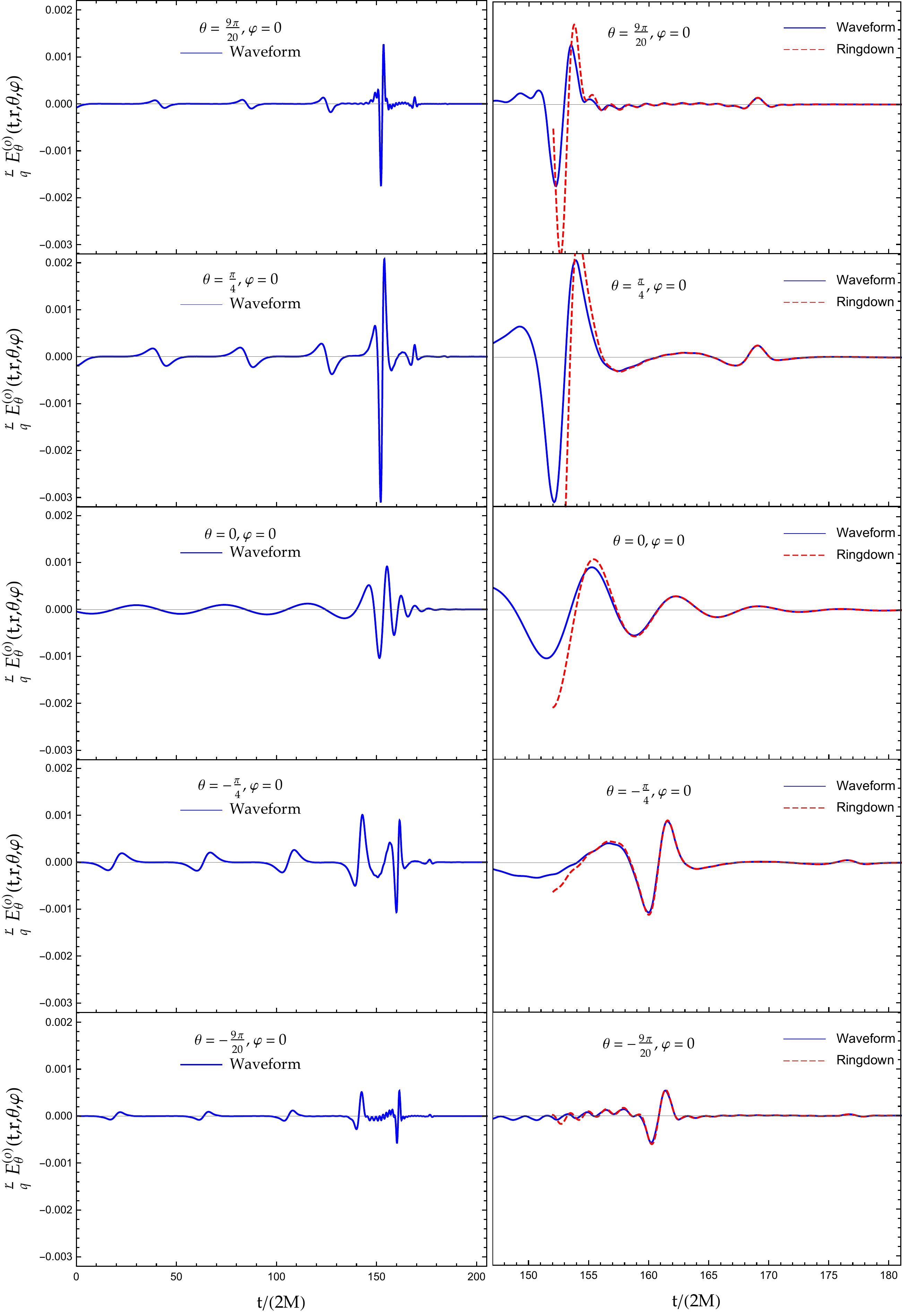}
\caption{\label{Etheta_Waveform_Vs_QNM_odd} Electric field $E_\theta^{(o)}$ observed at infinity for various directions above the orbital plane of the plunging particle. We consider $\varphi=0$ and we study the distortion of the multipolar waveform and of the associated quasinormal ringdown when $\theta$ varies between $-\pi/2$  and $+\pi/2$. We note that $E_\theta^{(o)}$ vanishes for $\theta = \pm \pi/2$ and that, for $\theta=0$, only the $(\ell=2,m=\pm 1)$ modes contribute to the signal.}
\end{figure*}

\begin{figure*}
\centering
 \includegraphics[scale=0.55]{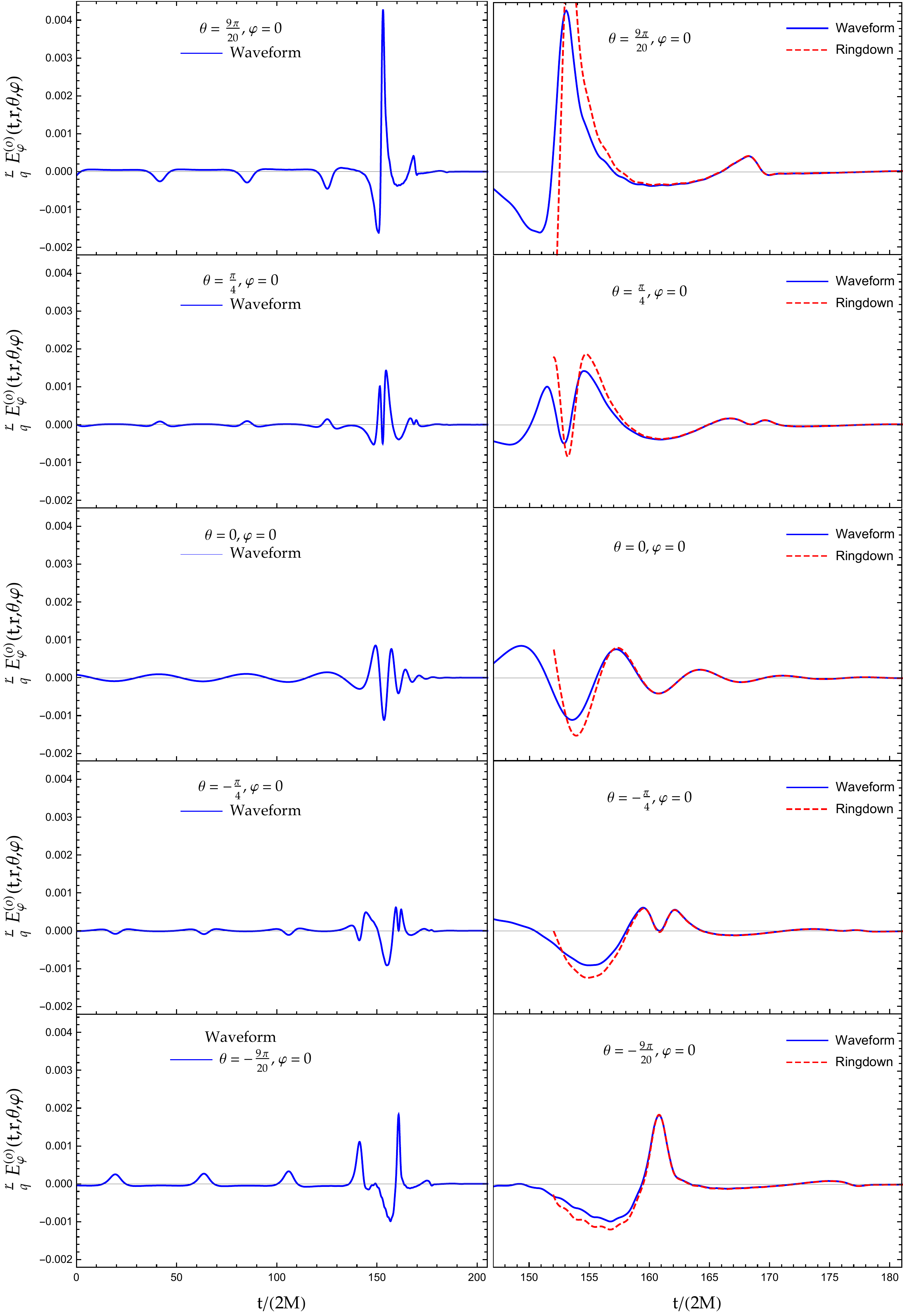}
\caption{\label{Ephi_Waveform_Vs_QNM_odd} Electric field $E_\varphi^{(o)}$ observed at infinity for various directions above the orbital plane of the plunging particle. We consider $\varphi=0$ and we study the distortion of the multipolar waveform and of the associated quasinormal ringdown when $\theta$ varies between $-\pi/2$  and $+\pi/2$. We note that, for $\theta=0$, only the $(\ell=2,m=\pm 1)$ modes contribute to the signal.}
\end{figure*}

\begin{figure*}
\centering
 \includegraphics[scale=0.53]{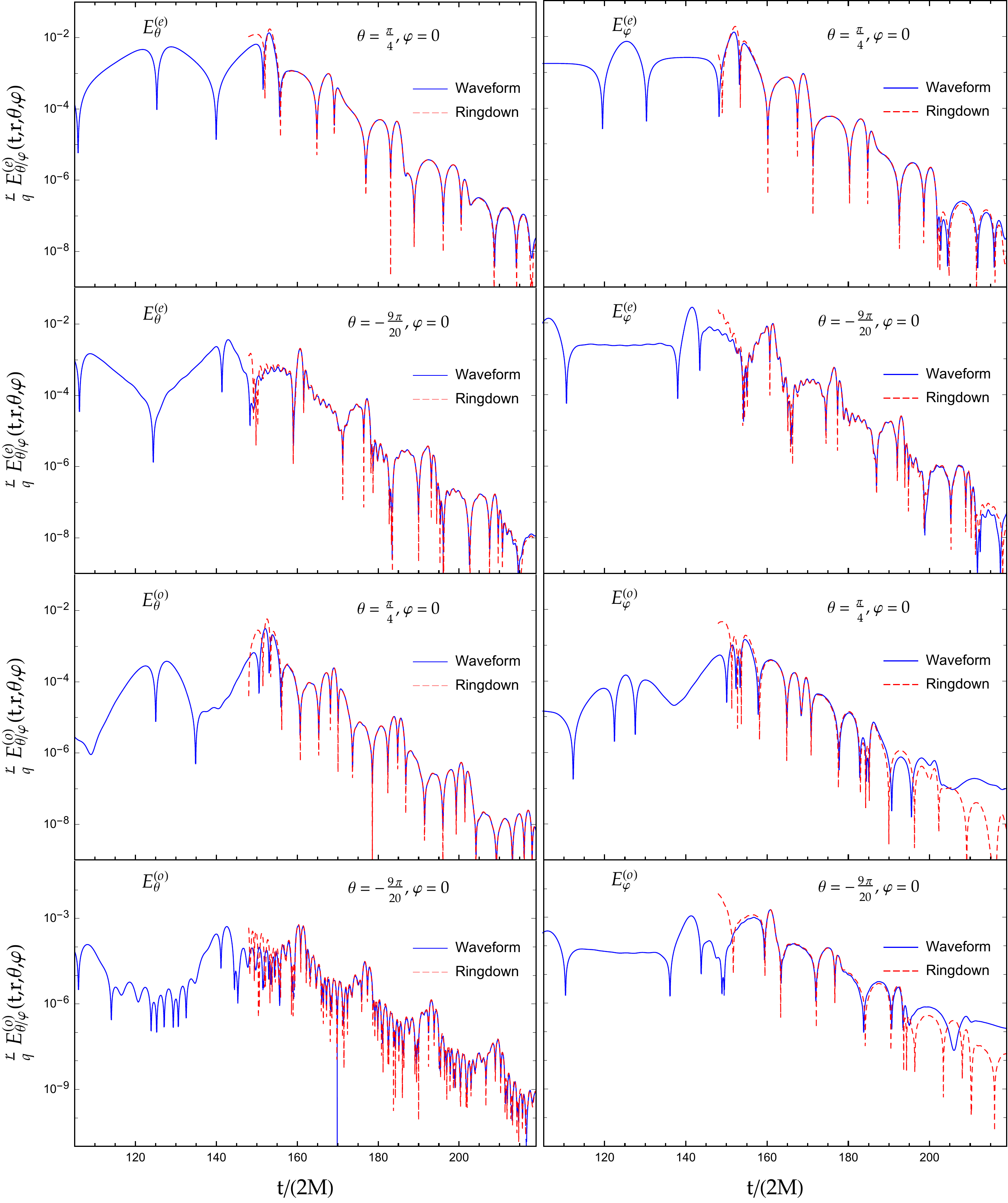}
\caption{\label{E_Waveform_Vs_QNM_log} Semi-log graphs of some multipolar waveforms showing the dominance of the quasinormal ringing at intermediate times and the agreement of the regularized waveforms with the unregularized quasinormal responses.}
\end{figure*}

\section{Multipolar waveforms and quasinormal ringdowns}
\label{SecIV}

\subsection{Numerical methods}
\label{SecIVa}

In order to construct the electric fields (\ref{ChampE_even}) and (\ref{ChampE_odd}), it is first necessary to obtain numerically the partial amplitudes $\psi^{(e/o)}_{\ell m}(t,r)$ given by (\ref{partial_response_def}). For that purpose, using {\it Mathematica} \cite{Mathematica}:

\begin{enumerate}[label=(\arabic*)]
   \item We have determined the functions  $\phi^{\mathrm{in}}_{\omega \ell}$ as well as the coefficients $A^{(-)}_\ell (\omega)$. This has been achieved by integrating numerically the homogeneous Regge-Wheeler equation (\ref{H_RW_equation}) with the Runge-Kutta method.  We have initialized the process with Taylor series expansions converging near the horizon and we have compared the solutions to asymptotic expansions with ingoing and outgoing behavior at spatial infinity that we have decoded by Pad\'e summation.

   \item  We have regularized the partial amplitudes $\psi^{(e/o)}_{\omega \ell m}(r)$ given by (\ref{Partial_Response_1}), i.e., the Fourier transform of the partial amplitudes $\psi^{(e/o)}_{\ell m}(t,r)$. Indeed, these amplitudes as integrals over the radial Schwarzschild coordinate are strongly divergent near the ISCO. This is due to the behavior of the sources (\ref{SourceTR_even}) and (\ref{SourceTR_odd}) in the limit $r \to 6M$. The regularization process is described in the Appendix. It consists in replacing the partial amplitudes (\ref{Partial_Response_1}) by their counterparts (\ref{REG_RES}) and to evaluate the result by using Levin's algorithm \cite{Levin1996}.

      \item We have Fourier transformed $\psi^{(e/o)}_{\omega \ell m}(r)$ to get the final result.

\end{enumerate}

\noindent Then, from the partial amplitudes $\psi^{(e/o)}_{\ell m}(t,r)$, it is possible to obtain the components $E_{\theta}^{(e/o)}$ and $E_{\varphi}^{(e/o)}$ of the electric field by using the superpositions (\ref{ChampE_even}) and (\ref{ChampE_odd}). We have constructed the even components from the $(\ell,m)$ modes with $\ell=1,\dots,10$ and $m=\pm \ell$ which constitute the main contributions. Similarly, we have constructed the odd components from the $(\ell,m)$ modes with $\ell=1,\dots,10$ and $m=\pm (\ell -1)$. In fact, it is not necessary to take higher values for $\ell$ because, in general, they do not really modify the numerical sums (\ref{ChampE_even}) and (\ref{ChampE_odd}).

In order to construct the quasinormal ringings associated with the electric fields (\ref{ChampE_even}) and (\ref{ChampE_odd}), it is necessary to obtain numerically the partial amplitudes $\psi^{\text{\tiny{QNM}} \, (e/o)}_{\ell m}(t,r)$ given by (\ref{partial_response_QNM_1}) and, as a consequence, the quasinormal frequencies $\omega_{\ell n}$, the excitation factors ${\cal{B}}_{\ell n}$ as well as the excitation coefficients ${\cal{C}}^{(e/o)}_{\ell m n}$ and ${\cal{D}}^{(e/o)}_{\ell m n}$. The quasinormal frequencies $\omega_{\ell n}$ can be determined by using the method developed by Leaver \cite{Leaver:1985ax}. We have implemented numerically this method by using the Hill determinant approach of Majumdar and Panchapakesan \cite{mp}. The excitation coefficients ${\cal{C}}^{(e/o)}_{\ell m n}$ and ${\cal{D}}^{(e/o)}_{\ell m n}$ can be ``easily'' calculated. Indeed, we first obtain the excitation factors ${\cal B}_{\ell n}$, as well as the functions $\phi_{\omega_{\ell n} \ell}^\mathrm {in}(r)$ and the coefficients ${A_{\ell}^{(+)}(\omega_{\ell n})}$ by integrating numerically the homogeneous Regge-Wheeler equation (\ref{H_RW_equation}) (for $\omega=\omega_{\ell n}$) with the Runge-Kutta method and then by comparing the solutions to asymptotic expansions with ingoing and outgoing behavior at spatial infinity. The evaluation of the integrals in Eqs.~(\ref{excitation_coeff_C}) and (\ref{excitation_coeff_D}) is then rather elementary because we do not have to regularize them. It should be noted that, for a given $\ell$, it is possible to consider only the fundamental QNM ($n=1$) which is the least damped one. Moreover, we need only the excitation coefficients ${\cal{C}}^{(e)}_{\ell m n}$ and ${\cal{D}}^{(e)}_{\ell m n}$ with $\ell=1,\dots,10$ and $m=\pm \ell$ and the excitation coefficients ${\cal{C}}^{(o)}_{\ell m n}$ and ${\cal{D}}^{(o)}_{\ell m n}$ with $\ell=1,\dots,10$ and $m=\pm (\ell -1)$. In Tables \ref{tab:table1} and \ref{tab:table2}, we provide the various ingredients permitting us to construct the quasinormal ringing associated with the electric fields (\ref{ChampE_even}) and (\ref{ChampE_odd}). It should be finally recalled that it is necessary to select a starting time $t_\mathrm{start}$ for the BH ringing. By taking $t_\mathrm{start}=t_p(3M)$, i.e., the moment the charged particle crosses the photon sphere, we have obtained physically relevant results.

\subsection{Results and comments}

In Figs.~\ref{E_Sum_ell}-\ref{E_Waveform_Vs_QNM_log}, we have considered the components $E_{\theta/\varphi}^{(e/o)}$ of the electric field  observed at infinity. The multipolar waveforms have been obtained by assuming that the particle starts at $r=r_\text{\tiny{ISCO}}(1-\epsilon)$ with $\epsilon=10^{-4}$ and, furthermore, in Eqs.~(\ref{trajectory_plung}) and (\ref{trajectory_plung_phi}), we have taken  $\varphi_{0}=0$ and chosen $t_{0}/(2M)$ in order to shift the interesting part of the signal in the window $t/(2M)\in[0,205]$. Without loss of generality, we have constructed only the signals for various directions above the orbital plane of the plunging particle. Indeed, we could obtain those observed below that plane by using the symmetry properties of the vector spherical harmonics in the antipodal transformation on the unit $2$-sphere $S^2$. Moreover, for the waveforms displayed in Figs.~\ref{Etheta_Waveform_Vs_QNM_even}-\ref{E_Waveform_Vs_QNM_log}, we have assumed that the observer lies in the plane $\varphi=0$. In fact, for any other value of $\varphi$, the behavior of the signals is very similar. The results corresponding to arbitrary values of $\theta$ and $\varphi$ are available to the interested reader upon request.

The distortion of the multipolar waveforms and of the associated quasinormal ringdowns appears clearly in Figs.~\ref{E_Sum_ell}-\ref{Ephi_Waveform_Vs_QNM_odd}. It can be observed in the ``adiabatic phase'' corresponding to the quasicircular motion of the particle near the ISCO (see Fig.~\ref{Trajectory_Plung}) as well as in the ringdown phase. It is due to the large number of $(\ell,m)$ modes considered in the sums (\ref{ChampE_even}) and (\ref{ChampE_odd}). In particular, the necessity to take into account a large number of modes to describe the waveforms is highlighted in Fig.~\ref{E_Sum_ell}. Moreover, it should be noted that the distortion of the signals is strongly dependent on the direction of the observer.

The multipolar waveforms and the associated quasinormal ringdowns are in excellent agreement as can be seen in Figs.~\ref{Etheta_Waveform_Vs_QNM_even}-\ref{Ephi_Waveform_Vs_QNM_odd} or, more clearly, in Fig.~\ref{E_Waveform_Vs_QNM_log} where we work with semi-log graphs. Here, it is important to recall (see Sec.~\ref{SecIVa}) that it has been necessary to regularize the former while the latter are unregularized.

\section{Conclusion and perspectives}
\label{Conc}

In this article, we have described the electromagnetic radiation emitted by a charged particle plunging from slightly below the ISCO into a Schwarzschild BH. For this, we have constructed the associated multipolar electromagnetic waveforms and analyzed their late-stage ringdown phase in terms of QNMs. Our results have been obtained for arbitrary directions of observation and have permitted us to emphasize more particularly the impact of higher harmonics on the distortion of the waveforms. It is moreover interesting to note the excellent agreement between the ``exact'' waveforms we had to regularize and the quasinormal waveforms which have not required a similar treatment.

Our work could be interesting in the context of multimessenger astronomy but it can also be considered as a warm-up with in mind the multipolar description, in the framework of BH perturbations, of the gravitational radiation produced by a ``massive particle'' plunging from the ISCO into a Schwarzschild BH \cite{Folacci:2018cic}. In particular, our present careful analysis of the theoretical and numerical difficulties linked with the regularization of the waveforms will be very helpful to deal with this more complicated problem which is of fundamental interest in gravitational wave physics. Indeed, the plunge regime from the ISCO is the last phase of the evolution of a stellar mass object orbiting near a supermassive BH or it can be also used to describe the late-time evolution of a binary BH (see, e.g., Refs.~\cite{Buonanno:2000ef,Ori:2000zn,Campanelli:2006gf,Nagar:2006xv,Sperhake:2007gu,
Mino:2008at,Hadar:2009ip,Hadar:2011vj,dAmbrosi:2014llh,Decanini:2015yba,Price:2015gia}).  Therefore, in this context, a multipolar description of the gravitational signal will be necessary with the enhancement of the sensitivity of laser-interferometric gravitational wave detectors (see, e.g., Ref.~\cite{Cotesta:2018fcv} and references therein).

\begin{acknowledgments}

We gratefully acknowledge Thibault Damour for drawing,
some years ago, our attention to the
plunge regime in gravitational wave physics. We wish also to thank Yves Decanini and Julien Queva
for various discussions and Romain Franceschini for providing us with powerful computing resources.

\end{acknowledgments}

\appendix*

\section{Regularization of the partial waveform amplitudes (even and odd parity)}
\label{appen}

In this Appendix, we shall explain how to regularize the partial amplitudes $\psi_{\omega \ell m}^{(e/o)}$. Indeed, the exact waveforms (\ref{Partial_Response_1}) as integrals over the radial Schwarzschild coordinate are strongly divergent near the ISCO. This is due to the behavior of the sources (\ref{SourceTR_even}) and (\ref{SourceTR_odd}) in the limit $r \to 6M$.

\subsection{Notations}

To do so, we introduce some notations in order to ``facilitate'' the singularity handling. We write the expressions (\ref{Partial_Response_1}) in the form
\begin{equation}\label{Rep_Part_}
\psi_{\omega \ell m}^{(e/o)} (r)=  e^{i \omega r_*} \psi^{(e/o)}_{\ell m}(\omega)
\end{equation}
with
\begin{equation}
\label{Rep_Part_Omega}
\psi^{(e/o)}_{\ell m}(\omega) = \gamma^{(e/o)} \int_{2M}^{6M} dr\, \phi_{\omega\ell}^{\mathrm{in}}(r){\cal A}^{(e/o)}(r)e^{i \Phi(r)}.
\end{equation}
Here,
\begin{subequations}\label{Fact_Gamma}
\begin{eqnarray}
& & \gamma^{(e)} =  \frac{1}{2 i \omega A_\ell^{(-)}(\omega)} \frac{q}{\sqrt{2 \pi}}  \, [Y^{\ell m}(\pi/2,0)]^*, \label{Fact_Gamma_even} \\
& & \gamma^{(o)} =  \frac{1}{2 i \omega A_\ell^{(-)}(\omega)} \frac{q}{\sqrt{2 \pi}}  \, [X^{\ell m}_\varphi(\pi/2,0)]^*, \label{Fact_Gamma_odd}
\end{eqnarray}
\end{subequations}
and
\begin{equation}
\label{Phase_tot}
\Phi(r) =\omega t_p(r) - m \varphi_p(r)
\end{equation}
while
\begin{subequations}\label{Amplitude_tot}
\begin{eqnarray}
& & {\cal A}^{(e)}(r) = \frac{18\sqrt{2}M \sqrt{r}}{(6M-r)^{5/2}}+i m \frac{12\sqrt{6}M r}{(6M-r)^{3}}  \nonumber \\
&& \qquad\qquad\qquad\qquad  -i\omega \frac{9r(r^2+12M^2)}{(6M-r)^{3}}, \label{Amplitude_tot_even} \\
& & {\cal A}^{(o)}(r) = \frac{6\sqrt{3}M}{\sqrt{r}(6M-r)^{3/2}}. \label{Amplitude_tot_odd}
\end{eqnarray}
\end{subequations}

\subsection{Regularization of $\psi_{\omega \ell m}^{(o)} (r)$}

Let us begin with the regularization of $\psi_{\omega \ell m}^{(o)} (r)$. Here, it is important to note that the amplitude ${\cal A}^{(o)}(r)$ given by (\ref{Amplitude_tot_odd}) diverges as $1/(6M-r)^{3/2}$ in the limit $r \to 6M$ and that the phase $\Phi(r)$ behaves as $1/(6M-r)^{1/2}$ in the same limit [or, in other words, that its derivative diverges as the amplitude ${\cal A}^{(o)}(r)$] while $\phi_{\omega\ell}^{\mathrm{in}}(r)$ is regular. As a consequence, the integrand in (\ref{Rep_Part_Omega}) which defines $\psi^{(o)}_{\ell m}(\omega)$ is a particular kind of rapidly oscillatory function and the integral (\ref{Rep_Part_Omega}) can be automatically ``regularized'' due to the oscillations induced by the phase term. In order to achieve the numerical neutralization of the divergence due to the amplitude ${\cal A}^{(o)}(r)$ by the oscillations induced by the phase $\Phi(r)$, we have used Levin's algorithm \cite{Levin1996} which is implemented in {\it Mathematica} \cite{Mathematica} and that permits us to obtain the numerical results of Sec.~\ref{SecIV}.

\subsection{Regularization of $\psi_{\omega \ell m}^{(e)} (r)$}

We can now consider the regularization of $\psi_{\omega \ell m}^{(e)} (r)$ which is much more complicated because the amplitude ${\cal A}^{(e)}(r)$ given by (\ref{Amplitude_tot_even}) diverges as $1/(6M-r)^3$ in the limit $r \to 6M$. However, we will be able to do it by reducing the degree of divergence of this amplitude from a series of integrations by parts of the integral (\ref{Rep_Part_Omega}) defining
$\psi^{(e)}_{\ell m}(\omega)$ and by dropping the boundary terms at $r=6M$ systematically. In doing so, we use a regularization process which is common in the context of gravitational wave physics \cite{Detweiler:1979xr}. But, here, to accomplish this task, we shall limit the number of successive integrations by parts by returning to an integral which can be likewise treated by Levin's algorithm.

To regularize $\psi_{\omega \ell m}^{(e)} (r)$, we need to split the amplitude ${\cal A}^{(e)}(r)$, the phase term $\Phi (r)$ and its ``derivative'' into a divergent and a regular part. As far as the amplitude ${\cal A}^{(e)}(r)$ is concerned, we write
\begin{equation}
\label{Amplitude_tot_bis}
{\cal A}^{(e)}(r)  = {\cal A}^{(e)}_{\mathrm{div}}(r) + {\cal A}^{(e)}_{\mathrm{reg}}(r)
\end{equation}
where the divergent part, which is obtained by the Taylor expansion of ${\cal A}^{(e)}(r)$ at $r = 6M$, is given by
 \begin{equation}
\label{Amplitude_div}
{\cal A}^{(e)}_{\mathrm{div}}(r)  =\frac{c_1}{(6M-r)^3} + \frac{c_2}{(6M-r)^{5/2}} + \frac{c_3}{(6M-r)^2}
\end{equation}
with
\begin{subequations}\label{Coeffs_ci}
\begin{eqnarray}
c_1& =& 18 i\, (2M)^2 \left[\sqrt{6}\, m - 36\, M\omega \right], \\
c_2 &=& 9\sqrt{6}\, (2M)^{3/2},\\
c_3 &=& 6 i\, (2M)  \left[-\sqrt{6}\, m + 90\, M\omega\right].
\end{eqnarray}
\end{subequations}
Here it should be noted that ${\cal A}^{(e)}_{\mathrm{reg}}(r)$ is not really the regular part of ${\cal A}^{(e)}(r)$ but, in fact, it is its part whose integral can be regularized by the oscillations of the phase $\Phi(r)$ using Levin's algorithm. Similarly, we also introduce the divergent and regular parts of the phase $\Phi(r)$ and of its derivative. We write
\begin{equation}
\label{Phase_tot_bis}
\Phi (r) = \Phi_{\mathrm{div}}(r) + \Phi_{\mathrm{reg}}(r)
\end{equation}
where
\begin{equation}
\label{Phase_div}
 \Phi_{\mathrm{div}}(r) = \frac{c}{\sqrt{6M-r}}
\end{equation}
with
\begin{equation}
\label{Coeff_c}
c =6\sqrt{2M}\left(m - 6\sqrt{6} \, M\omega \right)
\end{equation}
and we consider
\begin{equation}
\label{Theta}
\Theta(r) = \frac{d}{dr}\Phi_{\mathrm{reg}}(r)
\end{equation}
which we split as
\begin{equation}
\label{Theta_bis}
\Theta(r)  = \Theta_{\mathrm{div}}(r) + \Theta_{\mathrm{reg}}(r)
\end{equation}
where
\begin{equation}
\label{Theta_div}
\Theta_{\mathrm{div}}(r) = \frac{d}{\sqrt{6M-r}}
\end{equation}
with
\begin{equation}
\label{Coeff_d}
 d = \frac{m + 12\sqrt{6}  \, M\omega }{2\sqrt{2M}}.
\end{equation}
Of course, the coefficients $c$ given by (\ref{Coeff_c}) and $d$ given by (\ref{Coeff_d}) are obtained by Taylor expansions at $r = 6M$.

By inserting  (\ref{Amplitude_tot_bis}) into (\ref{Rep_Part_Omega}), we can write
\begin{equation}
\label{Rep_Part_Omega_bis}
\psi^{(e)}_{\ell m}(\omega) = \psi^{(e)\,{\mathrm{finite}}}_{\ell m}(\omega) + \psi^{(e)\,{\mathrm{div}}}_{\ell m}(\omega)
\end{equation}
where
\begin{subequations}\label{Psi_reg_div}
\begin{equation}
\label{Psi_reg}
 \psi^{(e) \, {\mathrm{finite}}}_{\ell m}(\omega)  = \gamma^{(e)} \int_{2M}^{6M} dr\, \phi_{\omega\ell}^{\mathrm{in}}(r){\cal A}^{(e)}_{\mathrm{reg}}(r)e^{i \Phi(r)}
\end{equation}
is finite (or, more precisely, can be regularized by Levin's algorithm) and where $\psi_{\ell m}^{(e) \, {\mathrm{div}}}(\omega)$, which is given by
\begin{equation}  \label{Psi_div}
\psi_{\ell m}^{(e) \, {\mathrm{div}}}(\omega) = \gamma^{(e)} \left[ {c}_1 I(3)+  {c}_2 I(5/2)+ {c}_3 I(2) \right]
\end{equation}
\end{subequations}
with
\begin{equation}
\label{I_generale}
I(\alpha) = \int_{2M}^{6M} dr\,\phi_{\omega \ell}^{\mathrm{in}}(r)\frac{e^{i \Phi(r)}}{(6M-r)^{\alpha}},
\end{equation}
is the sum of three divergent integrals. Now, we can reduce the degree of divergence of these integrals by integrating by parts $I(\alpha)$. We first note that [see Eqs.~(\ref{Phase_tot_bis}) and (\ref{Phase_div})]
\begin{eqnarray}
\label{I_generale_1}
I(\alpha) &= & \int_{2M}^{6M} dr\, \left(\frac{\phi_{\omega \ell}^{\mathrm{in}}(r) e^{i \Phi_{\mathrm{reg}}(r)}}{(6M-r)^{\alpha-3/2}}\right)\left(\frac{e^{i \Phi_{\mathrm{div}}(r)}}{(6M-r)^{3/2}} \right)\nonumber \\
               &=&\frac{2}{i c} \int_{2M}^{6M} dr\, \left(\frac{\phi_{\omega \ell}^{\mathrm{in}}(r) e^{i \Phi_{\mathrm{reg}}(r)}}{(6M-r)^{\alpha-3/2}}\right)\frac{d}{dr}\left(e^{i \Phi_{\mathrm{div}}(r)} \right) \nonumber\\
\end{eqnarray}
and then, taking into account Eqs.~(\ref{Theta}), (\ref{Theta_bis}) and (\ref{Theta_div}), and dropping intentionally the boundary term at $r = 6M$ (regularization), we obtain
\begin{widetext}
\begin{eqnarray}
\label{I_generale_2}
I(\alpha) = \frac{2 i}{c} \frac{\phi_{\omega\ell}^{\mathrm{in}}(2M)}{(4M)^{\alpha-3/2}} e^{i \Phi(2M)} &+& \frac{2 i}{c}\left(\alpha-\frac{3}{2}\right) I(\alpha-1/2)-\frac{2 d}{c} I(\alpha-1) \nonumber\\ &+&  \frac{2i}{c}\int_{2M}^{6M}dr\,\frac{\frac{d}{dr}\phi_{\omega \ell}^{\mathrm{in}}(r)}{(6M-r)^{\alpha-3/2}}\,e^{i\Phi(r)}-\frac{2}{c}\int_{2M}^{6M}dr\, \frac{\phi_{\omega\ell}^{\mathrm{in}}(r)\,\Theta_{\mathrm{reg}}(r)}{(6M-r)^{\alpha-3/2}}\, e^{i\Phi(r)}.
\end{eqnarray}
Now, by using recursively (\ref{I_generale_2}), we can extract from $\psi_{\ell m}^{(e) \, {\mathrm{div}}}(\omega)$ a finite contribution and we can finally replace the divergent function $\psi_{\ell m}^{(e)}(\omega)$ defined by (\ref{Rep_Part_Omega_bis}) and (\ref{Psi_reg_div}) by its regularized counterpart

\begin{eqnarray}\label{REG_RES}
& & \psi^{(e) \, \mathrm{reg}}_{\ell m}(\omega)= \gamma^{(e)} \int_{2M}^{6M} dr\, {\phi}_{\omega\ell}^{\mathrm {in}}(r) {\cal A}^{(e)}_{\mathrm{reg}}(r) e^{i \Phi(r)} \nonumber \\
& & \qquad\qquad\qquad + \frac{3\sqrt{6}}{2}\sqrt{2M}\gamma^{(e)}   \left[\int _{2M}^{6M}dr\, \phi_{\omega \ell}^{\mathrm {in}}(r) \left(\frac{1}{(6M-r)^{3/2}}+\frac{2id}{(6M-r)}\right) e^{i\Phi(r)}  \right.  \nonumber \\
& & \qquad\qquad\qquad\qquad\qquad\quad \left. -2 i\int_{2M}^{6M}dr\, rf(r)\, \frac{\phi_{\omega \ell}^{\mathrm {in}} (r)\, \Theta_{\mathrm{reg}}(r) }{(6M-r)^{3/2}} e^{i\Phi(r)} -2 \int_{2M}^{6M} dr\, rf(r) \frac{ \frac{d}{dr} \phi_{\omega \ell}^{\mathrm {in}}(r)  }{(6M-r)^{3/2}}e^{i\Phi(r)} \right].
\end{eqnarray}

\end{widetext}
Here, it is interesting to remark that, in Eq.~(\ref{REG_RES}), the boundary term at the horizon [i.e., a term analogous to the first term in the r.h.s.~of (\ref{I_generale_2})] has disappeared ``miraculously''! In fact, this is due to the cancellation of the coefficient in front of this boundary term which comes from the particular expressions of the coefficients $c_1$, $c_2$, $c_3$, $c$ and $d$ given by (\ref{Coeffs_ci}), (\ref{Coeff_c}) and (\ref{Coeff_d}).

\bibliography{EM_en_S}

\end{document}